# Topological Regard to Graphene: Elucidating the Morphology-Strain Correlation

Hadi Arjmandi-Tash, Alexander Kloosterman and Grégory F. Schneider[†]

*Faculty of Science, Leiden Institute of Chemistry, Leiden University, Einsteinweg 55, 2333CC Leiden, The Netherlands*

[†]to whom correspondence should be addressed: g.f.schneider@chem.leidenuniv.nl

**Abstract**

Graphene, dubbed as a two-dimensional material represents the topological concept of "surface" embedded in a three-dimensional space. This regard enables to employ existing theories/tools in topology to understand different properties/observations in graphene. Under the light of the long-established "Gauss's Theorema Egregium" we study wrinkled graphene, observing a peculiar correlation between morphology and strain distribution. Compressing graphene on water serves as an effectual platform to realize wrinkles; we explain the evolution of the wrinkles and the global distribution of the strain field while progressing the compression. The introduced platform in this paper offers an efficient approach to precisely control the generation and evolution of the wrinkles, transforming into a naturally occurring 3D landscape as a result of graphene buckling.

**Introduction**

Graphene is typically referred to as a strictly "two-dimensional material" [1]; Any material are composed of matters, thus inevitably possesses a finite thickness in the third dimension. Indeed the extension into the third dimension overrides the definition of the true two-dimensional topological object, the "surface" expressed as a zero-thickness entity [2]. While the surface is an abstract object, the two-dimensional material is physically accessible. The conceptions of a surface and a two-dimensional material, however could be mutually substituted in different purposes/applications to provide new insights/capabilities. In our earlier work, we have replaced graphene with a surface to benefit from its zero physical thickness in boosting the sensing capability of otherwise a graphene biosensor [3]. The surface was defined at the interface of two three-dimensional objects. Here in this work, on contrary, we replace a surface with a graphene, physically accessible to probe the correlation between strain and morphology, as explained by a known mathematical thorium for surfaces.

The Langmuir- Blodgett (LB) technique offers several potentials to study the physics of graphene. The ultra-flat surface of water host the graphene in its natural form avoiding the inclusion of any substrate-related perturbations normally observable on solid substrates [4]. The well-developed platform , on the other hand, is capable to apply an evolving compressive pressure on the film (graphene in this case) at air/water interface in a controllable fashion while the Wilhelmy technique provides a real-time feedback to correct the pressure [5]. The absence of any dry friction with the substrate (water) allows graphene to freely deform in response to the external compression. In our earlier report, we have qualified the LB technique in characterizing the mechanics of centimetre-scale graphene on water [6]. We further develop that study in this report with focusing on *post-mortem* analyses of graphene samples transferred on solid substrates. We disclose the evolution of the wrinkles at different stages of the compression.

**Experimental**

Preparation of the samples follows the approach we recently have developed[6,7]. Briefly, we placed chemically grown graphene on copper samples[8] (size = 1 × 1 cm$^2$) at air/liquid interface in a Langmuir Blodgett trough and surround the sample with a suitable amount of Dipalmitoylphosphatidylcholine (DPPC) lipid. The sub-phase − 500 mM of Ammonium persulfate solution in water − gradually etches away the copper foil and turns the bare graphene eventually floating on the surface (Figure 1-a, top panel). At this stage, the two barriers of the trough gradually move forward, pushing the lipids and applying isotropic bidirectional compressive pressures to graphene. We have shown[6] that in a progressive compression experiment, graphene samples initially exhibit an elastic phase where the deformation is reversible. Further compression of graphene beyond a limit buckles graphene and drives irreversible deformations, manifested by permanent out-of-plane wrinkles with tens of nanometers amplitude (Figure 1-b). Critical surface pressure at the onset of the elastic-to-plastic-deformation transition ($p_{cr}$) of our 20 samples approached ~30 $m$N/m[6]. We study the morphology and strain distribution in such wrinkles in this work. In practice, once the target surface pressure ($\geq$ 30 $m$N/m) is reached, we transfer the graphene on a SiOx/Si wafer by horizontally and gradually placing the wafer in contact with graphene by using the lift assembly of Langmuir–Schaeffer mode of the set-up (Figure 1-a, bottom panel). Graphene on a solid substrate is more suited for conventional characterization methods.

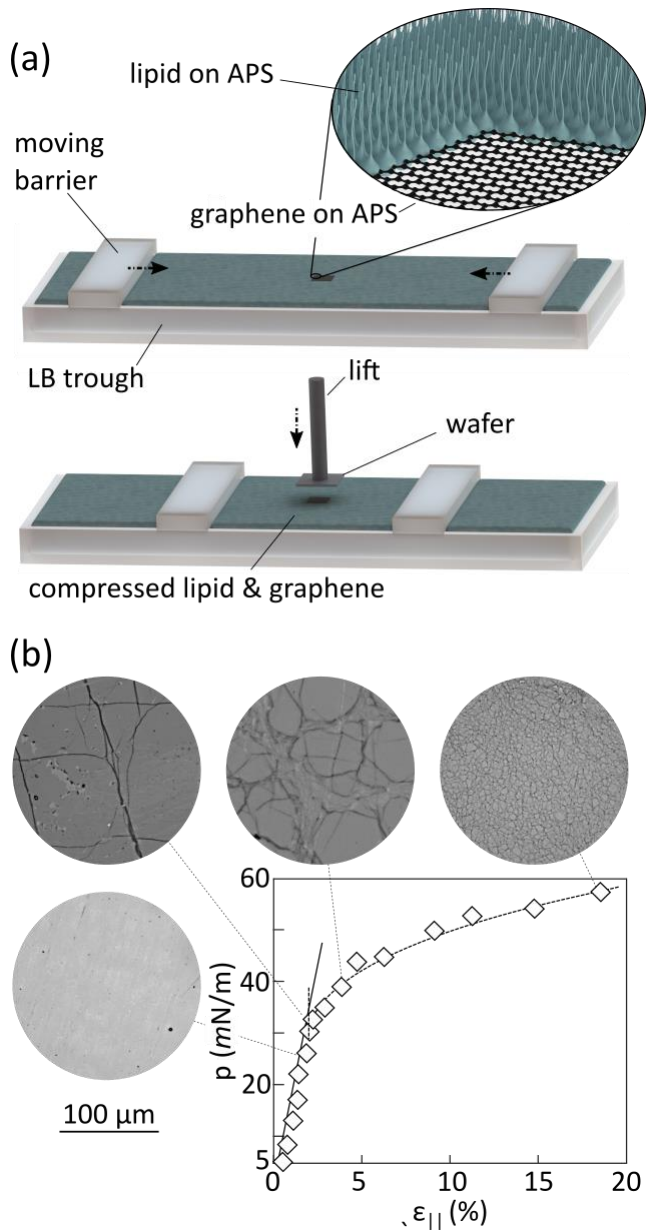

Figure 1: Langmuir Blodgett trough for preparation of crumpled graphene
a) Schematic representation of the set-up: Top) graphene is surrounded by lipids on a layer of APS solution. Bottom) Compressed graphene − achieved by lowering the area between the barriers − is transferred onto an oxidized silicon wafer for later characterizations. b) Typical stress (surface pressure)-strain plot of graphene samples, featuring elastic (linear) and plastic (nonlinear) deformation regions. Optical micrographs show the morphology of graphene samples at different compression levels.

**Figure 2** characterizes the optical micrographs of different graphene samples at the initial stages after buckling. Presence of the networks of (crossing) wrinkles is the most salient feature in these figures. Right after the buckling ($p \sim p_{cr}$, **Figure 2**-a) the wrinkles follow the principle axis of compressions (x and y directions) and are orthogonal to each other (T-shaped connections). Independent and subsequent (as oppose to simultaneous) formation of the x-aligned and y-aligned wrinkles best describes this pattern (see the inset schematic). Close to $p_{cr}$, the initially flat graphene first buckles along the direction which is "weaker", then then next generation of the wrinkles, perpendicular to the first pattern appear. Note that in application, the mechanical properties of centimetre-scale graphene is affected by the crystalline defects and initial static wrinkles; nothing guarantees identical resistance against buckling along both the x and y directions. Similar observation has been reported earlier in biaxial compression of elastic membranes before[9].

The samples further compressed up to p = 40 $m$N/m (slightly above $p_{cr}$) exhibit more complex network of the wrinkles (**Figure 2**-b): close to the edges of graphene, the force applied from the nearby edge is dominant (i: marginal zone); the stress is mainly uniaxial as many of the wrinkles are aligned parallel to the edge (**Figure 2**-c). Separation between the wrinkles may reach several micrometers, larger than the observed separations on polymeric substrates (less than 1 µm) [10]. Generation of a wrinkle demands a significant sliding of graphene over the substrate, accordingly the tribological properties of the graphene-substrate interface governs the separation between the wrinkles. Indeed large wrinkles with long wavelengths are energetically favorable in the absence of the friction [11], such as on water as observed here. Areas close to the centre (ii: central zone) of graphene exhibit a completely different morphology (**Figure 2**-d): Here, the deformations generate a landscape of mechanically self-assembled networks, enclosing nearly flat subdomains. The network does not exhibit any preferred direction indicating that the x and y components of the stress are comparable.

Further compression eliminates the featured distinct zones. Particularly when $p$ grows remarkably larger than $p_{cr}$, a dense directionless network of crossing wrinkles appears (Figure 3). The enclosed flat areas are now considerably smaller than in the central zone of p = 40 $m$N/m. The wrinkles are almost uniformly spread all over the sample. We note mechanically deforming graphene is an important approach to customize the performance of graphene-based devices [10,12]. Our study establishes a novel methodology to engineer the direction, the size and the density of the wrinkles in graphene.

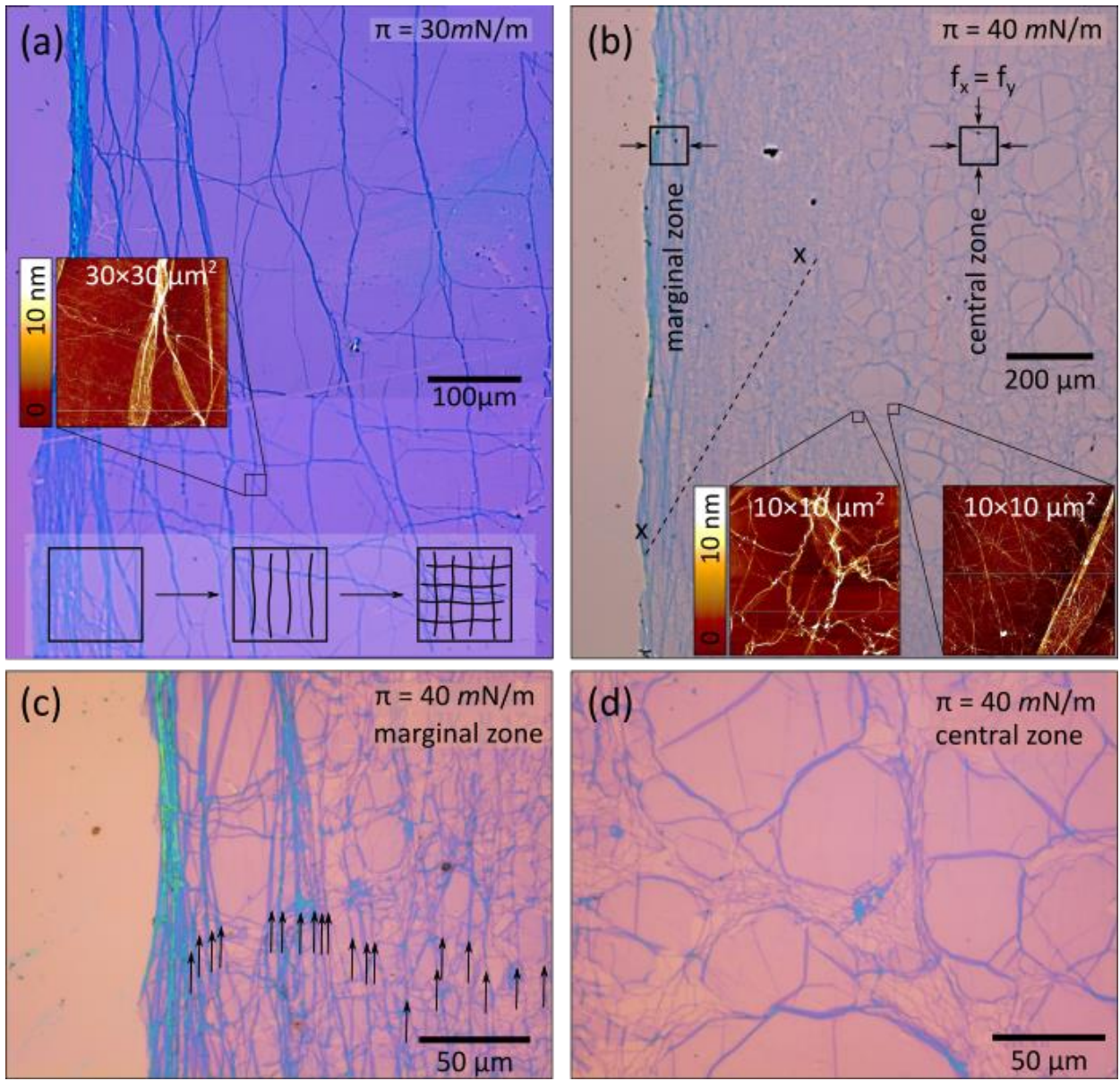

Figure 2: Morphological study of graphene samples close to buckling perturbation

a) Optical micrograph of a graphene sample transferred onto a SiO$_x$/Si wafer at p = 30 $m$N/m featuring a pattern of orthogonal wrinkles following the principal directions of anisotropy; inset AFM mapping probes a selected region on the sample. The bottom inset schematically demonstrates the generation of orthogonal wrinkles.

b) Optical micrograph of a graphene sample transferred onto a SiO$_x$/Si wafer at p = 40 $m$N/m: Dual marginal and central zones respectively featuring uniaxial and biaxial compressions are identified. inset AFM mappings probe selected regions on the sample.

c) Zoomed-in view to a selected window in the marginal zone of the same sample in b. Some wrinkles parallel to the border of the graphene are marked by arrows.

d) Zoomed-in view to a selected window in the central zone of the same sample in b.

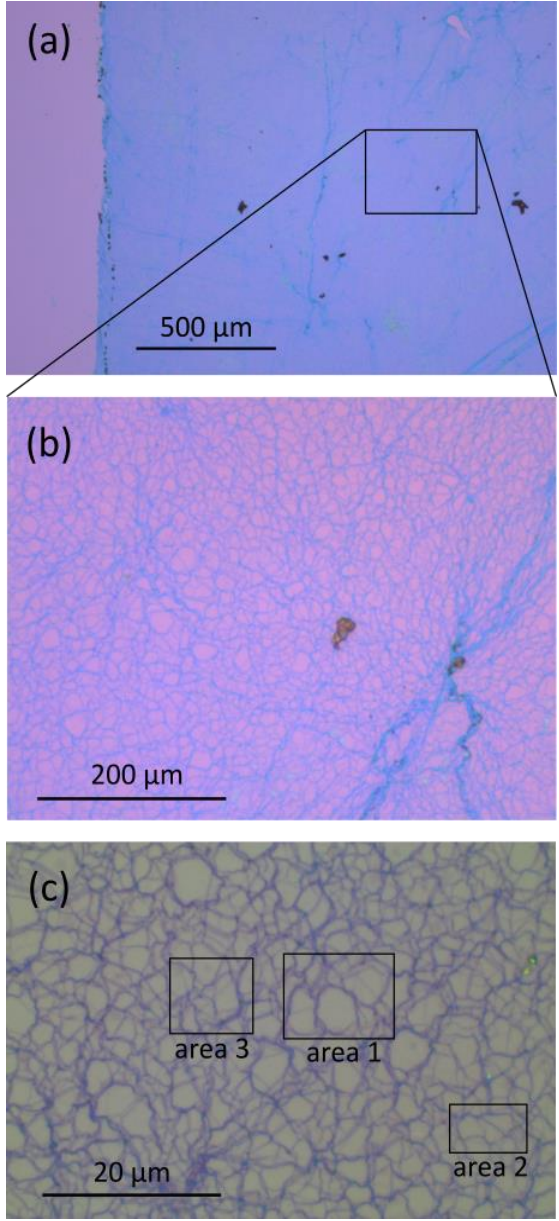

Figure 3: Morphological study of graphene samples at heavy compression regime (p >> p$_{cr}$)
a) Optical micrograph of a graphene sample transferred onto a SiO$_x$/Si wafer at p = 60 *m*N/m, featuring a pattern of crossing wrinkles, (semi) uniformly covering all over the sample
b) Magnified view to the window highlighted in a
c) Further magnified view to a selected window in a; the crossing networks of wrinkles enclose areas of few micrometer characteristic lengths.

The morphology of the wrinkles affects the distribution of the strain in graphene. **Figure 4** studies the correlation between the morphology and strain distribution in the triplet areas marked in Figure 3-c. The wrinkles and the enclosed flatlands are well-distinguishable in the AFM mappings. While some of the wrinkles are heavily crumpled and are of irregular shapes, some others are straight folds lied down on surface.

Raman spectroscopy provides an analytical tool to measure the strain in graphene; particularly it is confirmed that the position of the Raman 2D peak is sensitive to the strain deposited in graphene lattice [13–15]. Obviously the strain field is inhomogeneous and the strain is largely concentrated at heavily crumpled wrinkles where the 2D peak is upshifted (Figure 3 a III, bII and cIII). The flat areas, in contrast, exhibit a negligible strain as the 2D peak position averages to ~2675 cm$^{-1}$, a typical value for unstrained graphene [14,16–18]. The position of the 2D peak under compressive loadings exhibited a strain sensitivity of $\frac{\partial \omega_{||2D}}{\varepsilon} \sim 160 \ \left(cm^{-1}/\%\right)$ [18] which translates to $\frac{2693-2669\ (cm^{-1})}{160\ (cm^{-1}/\%)} = 0.15\%$ discrepancy between the peak and dip strains. The width of the 2D peak are always below 50 cm$^{-1}$ which is an standard value for monolayer graphene [19].

A careful examination of the strain distribution highlights the complex role of the morphology of deformations. Surprisingly, the straight wrinkles (some marked by ☆ in the AFM images) exhibit negligible contrast with the nearby graphene (background) in the Raman mapping. Those wrinkles are wide enough to be probed by Raman as even narrower but curly wrinkles were captured with the optics in the same mappings.

Gauss's Theorema Egregium assertion expresses the correlation between the morphology and strain distribution in surfaces [20]. Let's consider a flat surface as in **Error! Reference source not found.**-a, top. Numerous lines can be identified on this surface; they are all straight, i.e. of zero Gaussian curvature. By bending the surface in one direction (**Error! Reference source not found.**-a middle), some of the lines curl up (non-zero Gaussian curvature) while some others remain straight (zero Gaussian curvature). The Gaussian curvature of the hole surface is defined as the product of the minimum and maximum Gaussian curvature of the existing lines and remains zero. The assertion states that surfaces with identical Gaussian curvatures can transform to each other without straining. Bending the already curved surface in another direction (**Error! Reference source not found.**-a bottom) eliminates the straight lines, hence the hole surface gets a finite non-zero Gaussian curvature. Generation of such surfaces with double (multiple) curvatures from a flat surface − featured by the alteration of the Gaussian curvature − deviates the relative distances between arbitrary points, i.e. the straining of the surface. The same rationale is valid for graphene also (Figure 4). The straight wrinkles (marked by ☆ in the AFM images) share the same Gaussian curvatures of flat graphene from which they are originated. The formation of heavily crumpled wrinkles, however, alters the intrinsic Gaussian curvature of the flat graphene, involving straining. We note that the observation was already foreseen theoretically by atomistic simulation [10].

**Error! Reference source not found.**Figure 2-b provides a line map of the 2D peak position going from the marginal to central zones in mildly compressed graphene samples (see Figure 2-b). The marginal zone with dominant uniaxial compression, feature straight wrinkles (zero Gaussian curvature) while central zone exhibit characteristics of biaxial compression with heavily crumpled wrinkles (non-zero Gaussian curvature). The transition of the wrinkle morphology (as explained earlier) involves altering in strain level which is manifested by an upshift in 2D peak position. Note that in comparison to the uniaxial straining in which the formation of the wrinkles rapidly release the stretching energy, graphene under biaxial compression preserves the stretching energy even after wrinkle formation [21]. Stretching energy largely focuses on the wrinkle junctions.

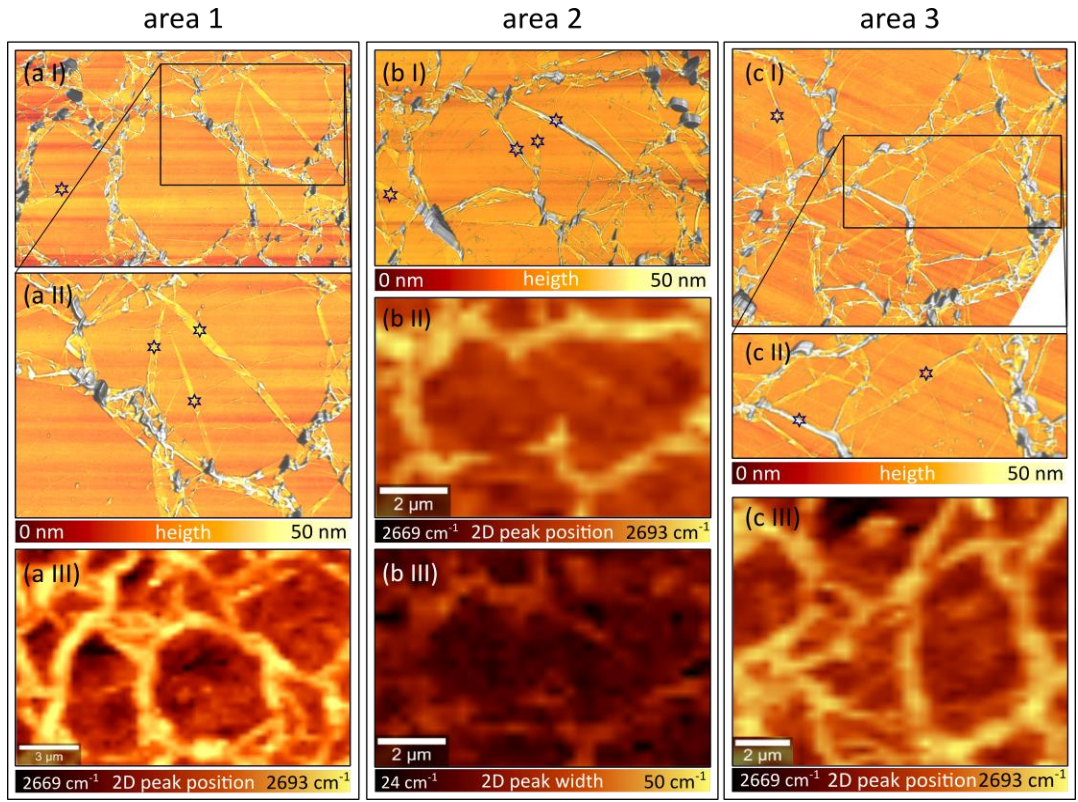

Figure 4: Strain distribution in crumpled graphene samples
a) Atomic force micrograph (I and II) and the Raman mapping of 2D peak position (III) corresponding to the window marked by "area 1" in Figure 3-c; AFM mappings features a network of heavily crumpled ridges in graphene while the Raman mapping disclose the strain distribution in the same area.
b) Atomic force micrograph (I) and the Raman mapping of 2D peak position (II) and full width at half maximum (III) corresponding to the window marked by "area 2" in Figure 3-c;
c) Atomic force micrograph (I and II) and the Raman mapping of 2D peak position (III) corresponding to the window marked by "area 3" in Figure 3-c;
All the Raman mappings were performed by using an excitation source of λ = 514 nm and power < 2mW. The utilized 100x objective provides a lateral resolution of ~ 300 nm for these mappings.

## Summary and Conclusion

The importance of wrinkled graphene stems from the fact that for a finite device size, the total surface of a wrinkled graphene exceeds − by far − a flat one. Indeed wrinkled graphene devices have found diverse applications ranging from detecting brain activities [22] to conductive coatings with controllable hydrophobicity [10] and to energy storage [23]. Successful fabrication approaches, however, are limited among which, buckling graphene on pre-strained flexible polymeric substrate is the most popular one [24]. As an important limitation, the substrate here is inappropriate for gating to build field effect transistors. Customizing wrinkles and strain level in graphene, on the other hand, requires precise calibration of strain in the polymeric substrate which is troublesome. Within this paper, we adopted LB trough to strain and graphene in a controllable fashion. Compressed graphene buckles beyond a certain limit, featuring large (~100s of nm) amplitude wrinkles. Negligible graphene/water interaction in this approach allows the graphene to deform freely. We studied the evolution of the wrinkles under the action of a progressive compression. Two dimensional graphene lattice with compression-induced wrinkles forms an standard system to probe the strain field where it couples to morphology, as explained by Gauss's Theorema Egregium. Our regard in graphene as a two-dimensional manifold is unprecedented in the literature.

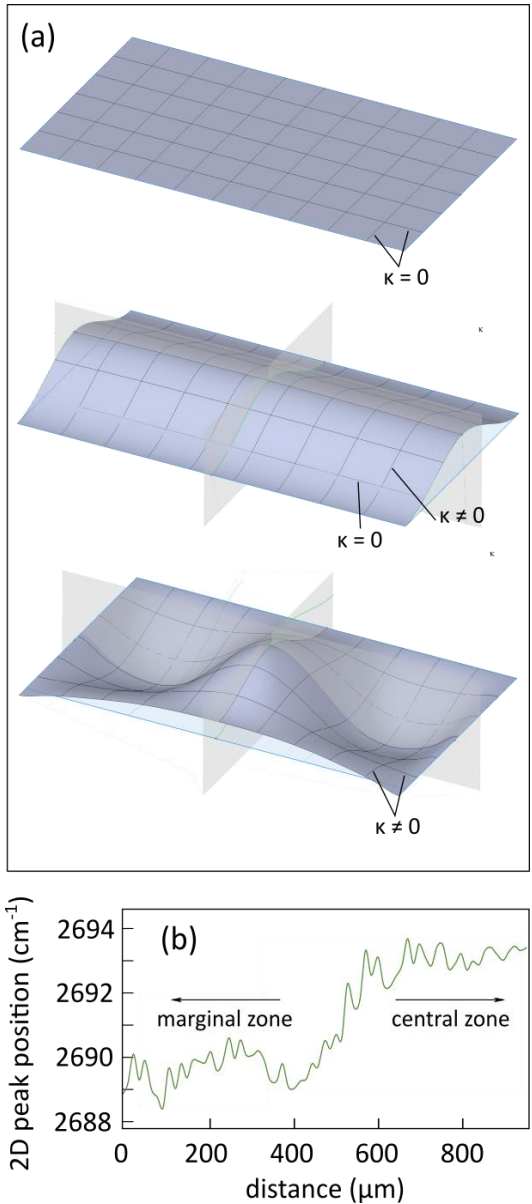

Figure 5: Gauss's Theorema Egregium to explain the corrolation of morphology and strain distribution in surfaces
a) Topological surfaces with no (top), mono (middle) and double (bottom) curvatures. κ corresponds to the Gaussian curvature of isolated arbitrary lines on corresponding surfaces.
b) Mapping of the Raman 2D peak position along the line marked by x-x on Figure 2-b. The Raman spectroscopy was performed with an excitation source of λ = 514 nm and power < 2mW. The utilized 10x objective averages over a spot size of ~ 5 μm to minimize the effect of local perturbations.

**Acknowledgements**

The work leading to this work has gratefully received funding from the European Research Council under the European Union's Seventh Framework Programme (FP/2007-2013)/ERC Grant Agreement No. 335879 project acronym 'Biographene', and the Netherlands Organisation for Scientific Research (Vidi 723.013.007). Khosrow Shakouri acknowledges financial support from the European Research Council through an ERC-2013 advanced grant (No. 338580). The authors thank Alexander Kros and Edgar Blokhuis for helpful discussions.